# DECONSTRUCTING THE QUANTUM DEBATE: TOWARD A NON-CLASSICAL EPISTEMOLOGY

Abstract

This article focuses on the problem of the authenticity of knowledge. It argues that the failure to integrate the observer into the act of observing is the main source of the disagreements and divisions in contemporary science and scholarship. The article uses the quantum debate as a case study to illustrate this point. It also argues that a truly non-classical epistemology should abandon the traditional subject-object orientation and focus on the very process of construction which creates both the subject and the object. The understanding of this process in terms of the equilibrium between equilibrium and disequilibrium will be critical for such epistemological perspective. The article concludes with some illustrations that show how the new perspective can help resolve the paradoxes which plague contemporary knowledge.

Authenticity of knowledge has been one of the most hotly debated problems in Western intellectual tradition throughout its history. The controversy, or rather multiple controversies, over the status of knowledge vis-à-vis reality go as far back as the Antiquity. They became particularly intense during the modern period. Kant's critique of the Enlightenment radically departed from the classical epistemology. Kant emphasized the constructed nature of knowledge and the role of human agency in knowledge production. His re-examination of the classical tradition led the way to a broad search for a non-classical epistemology that would integrate the new awareness and sensibilities.

Although this search has been going on for well over two centuries, no consensus seems to be emerging. On the contrary, debates over the status of knowledge have become increasingly more intense. Multiple sources generated this concern. It is beyond the scope of this paper to attempt a comprehensive historical account of its emergence and evolution. Suffice it to say that this increased concern about the status of knowledge reflects the growing awareness of autonomy and agency that has been broadly characteristic of the cultural climate that developed during the modern period, and particularly after the Second World War.[1] This awareness, more than anything else, catalyzed the debates over the status of knowledge.

These debates were part and parcel of practically every major development in Western thought during the 20[th] century. They accompanied the emergence of modern science and particularly quantum physics, were involved in the development of modern philosophical perspectives (such as post-structuralism and post-modernism), shaped new approaches in the study of our past (for example, the new historicism), brought innovations in literary criticism, and inspired, more recently, feminist and post-colonial critiques.

The intellectual perspectives that have transpired in these debates are very diverse. It would certainly be an oversimplification to reduce the entire array of unique and original contributions to a limited number of archetypal categories, such as representationalists and anti-representationalists or realists and anti-realists. The majority of contributors fall between such extremes. However, the fact that moderate perspectives in these debates exceed radical ones has had little effect on the intensity of the controversy. On the contrary, if anything, this intensity has increased, rather than subsided, as reputations continue to be created and destroyed, and the paradoxes generated in these debates continue to multiply. Moreover, there is hardly any sign that the controversy will be resolved any time soon. Its sustained nature indicates that cognitive disequilibria at the root of the divisions still endure and have yet to be addressed.

The current direction of these debates which focus on whether our knowledge corresponds to anything out there does not appear to be very constructive.[2] Even anti-representationalists agree that there are many instances in our experience when we do get things right and attain reliable knowledge that corresponds to the real world in which we live and function. So the question is not whether we can get things right, but rather how we get things right. How do we construct knowledge? What makes our knowledge reliable? Can we control knowledge production and make it more efficient?

Contemporary debates over these issues appear to be at an impasse and a genuine non-classical epistemology seems to be as distant a goal today as it was in Kant's lifetime. It is obvious that a resolution of the controversy and the construction of a genuine non-classical epistemology are not possible without a clear understanding of what produces this conundrum. This paper will explain the source of the controversy and sketch an outline of a non-classical epistemology.

**The Quantum Controversy**

There is hardly a case more suitable for an analysis of the controversy about authenticity of knowledge than the famous debate that took place over seventy years ago between Albert Einstein and Niels Bohr. Much has been written about this debate, which arguably was one of the most important and much discussed events in twentieth century physics. There is no need to add another detailed account to the host of excellent studies that have appeared over the years on the subject. A brief overview of the debate, however, may be in order.

The debate began in the late 1920s in the wake of the "second quantum revolution" when the so-called Copenhagen interpretation of quantum theory was formulated by its main proponents, among whom Niels Bohr and Werner Heisenberg were the most prominent contributors.[3] Einstein was certainly the most visible and celebrated figure among the critics of this interpretation. He and his supporters pursued several approaches in their criticism. They charged, for example, that quantum theory did not cover individual systems. Rather, they contended, it described the aggregate behavior of multiple systems, and not specific real situations. As Einstein put it in one of his contributions to this debate:

> What does not satisfy me in that theory, from the standpoint of principle, is its attitude toward that which appears to me to be the programmatic aim of all physics: the complete description of any (individual) real situation (as it supposedly exists irrespective of any act of observation or substantiation).[4]

Another line of criticism pursued by Einstein was to challenge the principle of complementarity. According to this principle, as articulated by Bohr, evidence obtained under different and mutually exclusive experimental conditions "cannot be comprehended within a single picture, but must be regarded as complementary. . . ."[5]

Einstein devised an ingenious experiment that became known as the photon-in-the-box experiment to provide an example which would demonstrate a possibility of measuring with arbitrary precision two complementary quantities, regarded as non-commuting, at the same time and thus disprove one of the fundamental principles of quantum theory.[6]

Still another important line of criticism—one that eventually appeared in the paper published in 1935 jointly by Einstein, Podolsky, and Rosen—alleged that the Copenhagen interpretation was inconsistent.[7] Einstein found it paradoxical that the measurement on one of the particles, which had been in contact and then became separated, would still affect the measurement of another particle—what Einstein called the "spooky action at a distance." Although Einstein repudiated the EPR paper soon after its publication,[8] he nevertheless continued to insist that "[t]he assertion that, in this latter case [i.e. separated systems], the real situation of B could not be (directly) influenced by any measurement taken on A is, therefore, within the framework of quantum theory unfounded and (as the paradox shows) unacceptable."[9] In a letter to Rosenfeld, Einstein asked, largely rhetorically: "How can the final state of the second particle be influenced by a measurement performed on the first, after all physical interaction has ceased between them?"[10]

All these specific criticisms converged on one major point that quantum theory did not provide a complete description of the microscopic processes and that physicists should continue searching for a more comprehensive approach.

The intense criticisms left the Copenhageners unimpressed. In response to charges that quantum theory refused to deal with individual phenomena but only with aggregates, Bohr argued that quantum theory had a very different conception of individuality than did classical physics. "In fact," he wrote, ". . . in quantum physics we are presented . . . with the inability of the classical frame of concepts to comprise the peculiar feature of indivisibility, or 'individuality,' characterizing the elementary processes."[11] In his view, "the quantum postulate requires that 'interactions' between 'systems' be accorded the feature of 'wholeness' or 'individuality.'"[12] Bohr also countered Einstein's ingenious photon-in-the-box experiment with his own interpretations of this experiment, invoking Einstein's own theory of relativity to show that the experiment did not violate the principle of complementarity,[13] and moreover, it was well "suited to embrace the characteristic features of individuality of quantum phenomena."[14] The Copenhageners also did not appear to be perturbed by the paradoxes highlighted in the EPR paper. In fact, they did not see much new in this article that had not been addressed and explained by them prior to the paper's publication.[15] Bohr, for example, saw the argumentation advanced in the article as ill suited "to affect the soundness of quantum-mechanical description, which is based on a coherent mathematical formalism covering automatically any procedure of measurement like that indicated [in the article]."[16]

Although by the end of the 1930s the debate seemed to have passed its peak, Einstein persisted in his attempts to reason with his opponents. After failing to win the battle, Einstein tried to make a compromise. While granting that "the (testable) relations which are contained in it [quantum theory] are, within the natural limits fixed by the indeterminacy relations, <u>complete</u>,"[17] he argued that there was no ground, either theoretical or empirical, to consider any theory as a complete description of physical reality. He even conceded the incompleteness of his own theory of relativity, recognizing that "the general theory of relativity furnished . . . a field theory of gravitation, but no theory of the field-creating masses."[18]  But all these efforts were to no avail.  Bohr and his supporters remained unwavering in their stance and continued to gain new supporters for their orthodoxy.  Bohr managed to create a numerous cohort of loyal followers who revered him as a demi-god,[19] in part because of his remarkable achievements in physics but also, as Gell-Mann has observed, because he "brainwashed a whole generation of theorists into thinking that the job [an authentic and complete explanation of physical reality] was done fifty years ago."[20]

This visceral confrontation was not a mere clash of personalities or a competition for power—although there was definitely something of both in it.  One can explain its character by the fact that it touched upon some very fundamental problems concerning reality and its knowability.  These problems were at the time and continue to be to this day central for many physicists, and not only physicists.  For both Einstein and Bohr the issues on which they disagreed went to the very core of the meaning of the physical world that they constructed.  Einstein wrote in one of his letters to Schrodinger:  "The Talmudic philosopher [as he referred to Bohr] doesn't give a hoot for 'reality' which he regards as a hobgoblin of the naïve . . . ."[21]  To the end of his life Einstein remained convinced that ". . . to believe that [quantum mechanics is a complete account of reality] is logically possible, but it is so very contrary to my scientific instinct that I cannot forego the search for a more complete conception."[22]

Although the principal contributors to this debate have long been gone, the issues raised in this debate continue to be a source of controversy.  As Stent summarized:  ". . . among contemporary philosophers of science and philosophically inclined physicists, the basic epistemological problems raised by the photon-in-the-box experiment and the EPR

argument are not only still hotly debated, but their resolution is not even in sight.[23]

**The Epistemology of the Copenhagen Interpretation**

Two important elements are characteristic for the epistemological perspective of

the Copenhagen interpretation of quantum mechanics. One is a constant introduced by

Planck—the so-called Plank constant—to describe the size of quanta involved in exchange of energy on the subatomic level. The other is the recognition of the agency of the observer. The combination of these two features constitutes the core the Copenhagen interpretation of quantum mechanics.

The Copenhageners argued that classical physics, which viewed objects as

systems isolated from external influences, was totally inadequate for studying subatomic

particles. Due to the size and uncertainty involved in the behavior of quantum systems—

for example, they could exhibit features of a wave or a particle under different

experimental conditions—these systems could not be observed in isolation but only in

interaction with instruments used in the experiment. The energy in such interactions

could only be exchanged in fixed amounts called quanta—the smallest indivisible packets

of energy introduced in modern physics by Max Planck. Bohr and other Copenhageners

shared the view that ". . . an independent reality in the ordinary [that is, classical]

physical sense can neither be ascribed to the phenomena nor to the agency of observation."[24] However, while recognizing the indivisibility and distinctly non-classical character of quantum mechanical phenomena, they also stressed that there was no other alternative but to describe them in the classical mode, that is, differentiating between the behavior of a quantum mechanical system and the means of observing. Bohr tells us that ". . . it is decisive to recognize that, <u>however, far the phenomena transcend the scope of classical physical explanation, the account of all evidence must be expressed in classical terms.</u>[25]"

Thus, the epistemological foundation of the Copenhagen interpretation of quantum mechanics included both non-classical and classical components. One interesting consequence of this peculiar combination of a classical perspective, which postulated a gap between the subject and the object, and a non-classical approach, with its recognition of the agency of the observer, was that in the final analysis the epistemological perspective of quantum mechanics preserved classical dualism. One can certainly agree with Stapp who characterized the epistemology of the Copenhagen interpretation as a "hybrid" and a "half-way house"[26]: while it recognized the agency of the observer, it also preserved the gap between the observer and the observed.

The Copenhageners justified this division, which became known as the "Heisenberg cut," primarily by instrumental pragmatic considerations. While quantum mechanical processes involve uncertainty and ambiguity, the results must be presented unambiguously. Such unambiguous presentation requires that the object of observation be clearly differentiated from the observing instruments. But how can one achieve such differentiation given the quantum postulate? The answer of the Copenhageners was: by providing a description of the conditions of the experiment in terms of classical physics with its sharp differentiation between the subject and the object. As Bohr explained:

> The argument is simply that by the word <u>experiment</u> we refer to a situation where we can tell others what we have done and what we have learned and that, therefore, the account of the experimental arrangement and of the results of the observations must be expressed in unambiguous language with suitable application of the terminology of classical physics.[27]

Elsewhere Bohr further elucidated:

All description of experience so far has been based on the assumption, already inherent in ordinary conventions of language, that it is possible to distinguish sharply between the behaviour of the objects and the means of observation. This assumption is not only fully justified by everyday experience, but even constitutes the whole basis of classical physics.[28]

Some researchers, as for example, Landsman, see in these pragmatic justifications a keen concern for the objectivity of physics which calls for unambiguous descriptions:

It is precisely the objectivity of classical physics in this sense that guarantees the possibility of what Bohr calls "unambiguous communication" between observing subjects, provided this communication is performed in "classical terms."[29]

Others explain them as a mere inertia on the part of the Copenhageners that prevented them from completely abandoning classical conceptions. Stapp concludes:

The Copenhagen version of quantum theory is thus a hybrid of the old familiar classical theory, which physicists were understandably reluctant to abandon completely, and a totally new theory based on radically different concepts.[30]

Notwithstanding the pragmatic justifications enunciated by the framers of quantum theory, there are some indications that the reasons for the preservation of classical dualism go much deeper. Interaction between subatomic particles was the main focus of quantum theory. Bohr, Heisenberg, Pauli and others theorized this interaction in terms of energy exchange. The exchange of energy, according to quantum theory, is not a continuous process; it occurs, as Planck established, in the smallest fixed packets of energy called quanta or, as Bohr often called them, quanta of action.  In his explanation

of the connection between Planck's discovery and quantum theory, Bohr wrote:

> From the very beginning the main point under debate [with Einstein] has
> been the attitude to take to the departure from customary principles of
> natural philosophy characteristic of the novel development of physics
> which was initiated in the first year of this century by Planck's discovery
> of the universal quantum of action.  This discovery, which revealed a
> feature of atomicity in the laws of nature going far beyond the old doctrine
> of the limited divisibility of matter, has indeed taught us that the classical
> theories of physics are idealizations which can be unambiguously applied
> only in the limit where all actions involved are large compared with the
> quantum.[31]

The distinct feature of the quantum of action is its indivisibility:  we cannot differentiate independent behavior of a quantum system when it interacts with another quantum system or with a measuring device.  ". . .  [I]n quantum physics," Bohr tells us,

"we are presented  . . . with the inability of the classical frame of concepts to comprise

the peculiar feature of indivisibility, or 'individuality,' characterizing the elementary

processes."[32]  A crucial implication of the principle of indivisibility is that, in contrast to

classical physics which claims knowledge of objects as they are, that is, without in any

way disturbing the system, in quantum mechanics one cannot in principle know anything

about a quantum system as it is, outside the experiment set up by the observer.  In the

course of his debates with Einstein, Bohr stresses repeatedly that ". . . in quantum

mechanics, we are not dealing with an arbitrary renunciation of a more detailed analysis

of atomic phenomena, but with a recognition that such an analysis is in principle

excluded."[33]

This and similar statements show that although the immediate preoccupation of the Copenhagen interpretation was subatomic particles, its conclusions reached way beyond physics.  Implicit in this interpretation was a more far ranging philosophical claim about the knowability of reality. The Copenhageners posited quantum interactions as an ontological reality which set the absolute limit to what could be known about reality. It was to the setting of such limitation that Einstein reacted most strongly in his contributions to this debate. One can certainly agree with Landsman who astutely observes that ". . . the technical parts of their [Bohr and Einstein] debate on the (in)completeness of quantum mechanics just served as a pale reflection of a much deeper philosophical disagreement between Bohr and Einstein about the knowability of nature."[34]

The theory of knowledge that the Copenhageners deployed in their version of quantum mechanics is rooted in the atomistic tradition that goes back to Leucippus and Democritus. The main precept of this tradition with its analytical methodology is that knowledge can be attained by breaking complex entities into simple ones. For the Copenhageners, quantum theory reached the absolute limit of the atomistic approach in studying reality. It was the end of the line where, in their view, analysis and synthesis came together. Bohr explained:

> The question at issue has been whether the renunciation of a causal model
> of description of atomic processes involved in the endeavours [sic!] to
> cope with the situation should be regarded as a temporary departure from
> ideals to be ultimately revived or whether we are faced with an irrevocable
> step towards obtaining the proper harmony between analysis and synthesis
> of physical phenomena.[35]

Although theory of knowledge played such an important part in the Copenhagen interpretation, its authors did not make an attempt to base it on a more solid foundation than their intuitions. Their contributions to the debate do not contain any critical reflections on the theory of knowledge they deployed. They show no familiarity with both theoretical and empirical studies on how knowledge is constructed, even though such studies were available at the time, for example through works of Jean Piaget.[36] Finally, they do not provide any independent justifications for the limits they were setting up for human knowledge in their quantum theory. Bohr, for example, used the principle of complementarity which followed from the quantum principle to argue in support of setting such limitation:

> In fact, the individuality of the typical quantum effects finds its proper expression in the circumstance that any attempt of subdividing the phenomena will demand a change in the experimental arrangement introducing new possibilities of interactions between objects and measuring instruments which in principle cannot be controlled.[37]

In other words, he made a circular argument in which deduction justified the premises from which it was made. This fact did not escape Einstein who, in his "Reply to Criticism," after making a long argument that the statistical quantum theory allowed only descriptions of ensembles and not individual systems,[38] made an astute remark about circularity which made "hidden use of the object to be defined."[39]

The above shows that the inclusion of the classical component in quantum theory was not a mere pragmatic convenience for the Copenhageners. In formulating their theory, they came face to face with the problem of subjectivity and the construction of knowledge. They had no idea how to resolve this problem and they looked away. Their insistence on the classical description of the conditions of an experiment was hardly a solution to the problem of subjectivity; it merely shifted this problem to another level. If the idea of quantum mechanics was to recognize subjectivity and thus establish control over it, it certainly failed to achieve this goal: with the classical description of the conditions of an experiment, subjectivity was still on the loose and uncontrolled.

The failure to integrate subjectivity could place in doubt the entire enterprise of quantum mechanics to provide an objective description of physical reality. For the Copenhageners, this outcome was certainly unacceptable. Their way out of this predicament was simply to declare the ontological reality of quantum systems inaccessible to knowledge, thus in fact reviving the Kantian thing-in-itself but on a new foundation and creating a new orthodoxy.

The classical tradition was largely oblivious to the agency of the observer. Mental constructions were projected unconsciously and uncritically on reality which was regarded as given.[40] Dualism was a necessary outcome of this perspective which did not recognize a vital and constructive link between the subject and the object. Obviously, if one does not understand the process which constructs both the subject and the object, the two appear as ontologically separate. The classical epistemology allowed only one path to knowledge: more or less passive reflection.

Kant made the first major step in changing the view on the role of the observer. However, while recognizing the agency of the observer, he did not see the actual physical link that connected the observer and the object. In his view, knowledge was essentially rooted in self-referentiality of the observer whose mental a priori categories mediated physical experience. As a result, observing appeared as an infinite self-referential regression and the object as a thing-in-itself which in principle was inaccessible to knowledge.

In contrast to Kant, the Copenhageners saw the actual physical link between the subject and the object, but for them this link was nothing more than an undifferentiated mystery of the quantum of action. As a result, the object of quantum reality—the quantum system—appeared to them as an inaccessible thing-in-itself, while observing became an infinite self-referential introspection. In his famous statement, Heisenberg summarized:

> The conception of the objective reality of the elementary particles has thus evaporated not into the cloud of some obscure new reality concept, but into the transparent clarity of mathematics that represents no longer the behavior of the particles but rather our knowledge of this behavior.[41]

Bohr echoed the same idea when he wrote: ". . . when searching for harmony in life one must never forget that in the drama of existence we are ourselves both actors and spectators."[42]

The implications of self-referentiality in quantum theory could not but cause controversy. As Wiseman cogently observes:

> Heisenberg by 1958 had been seduced by the idea that "the discontinuous change in our knowledge in the instant of registration . . . has its image in the discontinuous change of the probability function [$\Psi$]." But he failed to take the necessary logical step of accepting that $\Psi$ can represent only one individual's knowledge.[43]

The subjectification of knowledge by the Copenhageners, as evidenced in the above quotes from Heisenberg and Bohr, was fraught with inconsistencies and made quantum theory vulnerable to criticism. In order to protect the new orthodoxy, the Copenhageners had to build defenses from both internal instability and external criticisms. In an effort to forestall possible fragmentations, Bohr resorted to policing the language used in quantum mechanical descriptions. As he described himself:

> . . . I warned especially against phrases, often found in the physical
> literature, such as "disturbing of phenomena by observation" or "creating
> physical attributes to atomic objects by measurements." Such phrases,
> which may serve to remind of the apparent paradoxes in quantum theory,
> are at the same time apt to cause confusion, since words like "phenomena"
> and "observations," just as "attributes" and "measurements," are used in a
> way hardly compatible with common language and practical definition.[44]

It is not surprising that in the charged atmosphere of the acrimonious debate, the search for knowledge turned, as many have agreed since, into an exercise of power.[45] Some historians of science even see Bohr as a Stalin[46] who managed to suppress the opposition through "a combination of political maneuvering, shrewd rhetoric, and spellbinding his colleagues and the general audience by the allegedly unfathomable depth of his thoughts . . . ."[47] Gell-Mann characterizes the advocacy of their doctrine by the Copenhageners as nothing short of "brainwashing."[48]

Einstein's position in this debate was also perturbed by classical dualism. In his general and special theory of relativity, Einstein pioneered the modern view which also emphasized the agency of the observer. In his theoretical perspective, observations also depend on the frame of reference chosen by the observer. Just like the Copenhageners, Einstein also did not follow through on his innovation: the interaction between subject and object remained, for Einstein, also a mystery. As a result, his innovative theories, just like quantum mechanics, conserved classical dualism. For example, one of the main principles of Einstein's epistemology was Trennungsprinzip (separability principle). Don Howard explains that "Einstein regarded his separation principle . . . as virtually an axiom for any future fundamental physics."[49] He invoked this principle in his critique of Bohr's complementarity and sustained a profound lifelong philosophical commitment to it, which explains his enduring uneasiness about quantum mechanics.

Einstein's insistence on <u>Trennungprinzip</u> was based on a belief that an assumption about independent existence of physical systems in space and time was absolutely essential in studying physical reality.  As he explained in his letter to Born:

> . . . if one renounces the assumption that what is present in different parts of space has an independent, real existence, then I do not at all see what physics is supposed to describe.  For what is thought to be a "system" is, after all, just conventional, and I do not see how one is supposed to divide up the world objectively so that one can make statements about the parts.[50]

According to Don Howard, Einstein's enduring commitment to the separability principle reflects his early interest in the philosophy of Arthur Schopenhauer, whom Einstein read at the Zurich Polytechnic.  Schopenhauer's <u>principium individuationis</u>, conditioned by space and time, led Einstein to conclude that the difference in location of any two systems was sufficient for regarding physical systems as independent of each other.  By the same analogy, he also regarded the subject and the object as two separate systems which existed in space and time.  In Einstein's own words, "the belief in the external world independent of the perceiving subject is the basis of all natural sciences."[51]

Thus Einstein's conception of subject-object relations was not significantly different from that of the Copenhageners.  In fact, although Einstein was not as steeped in the Kantian tradition as the latter were, he acknowledged some fundamental similarities between his views and Kant's philosophy.  Just like Kant, he recognized the fundamental gap separating the thing-in-itself from the observer.  The only difference that he saw between himself and Kant in this respect was that, in contrast to Kant, he did not regard cognitive categories (which, in Kant's view, were essential for constructing knowledge about reality) as immutable <u>a priori</u> ideas, but rather as free and changeable conventions. In his own explanation of his theoretical perspective, Einstein wrote:

The theoretical attitude here advocated is distinct from that of Kant only by the fact that we do not conceive of the "categories" as unalterable (conditioned by the nature of the understanding) but as (in the logical sense) free conventions. They appear to be <u>a priori</u> only insofar as thinking without the positing of categories and of concepts in general would be as impossible as is breathing in a vacuum.[52]

Contrary to the criticism for allegedly adhering to the classical tradition directed against him by the Copenhageners, his idea of the real was not that much different from their thing-in-itself. Again, in Einstein's own words:

I did not grow up in the Kantian tradition, but came to understand the truly valuable which is to be found in his doctrine, alongside of errors which today are quite obvious, only quite late. It is contained in the sentence: "The real is not given to us, but put to us (<u>aufgegeben</u>) (by way of riddle)." This obviously means: There is such a thing as a conceptual construction for the grasping of the inter-personal, the authority of which lies purely in its validation. This conceptual construction refers precisely to the "real" (by definition), and every further question concerning the "nature of the real" appears empty.[53]

Einstein's advocacy of the "real" in his debates with Bohr was not due to the alleged classical prejudices, but rather, as in the case of the Copenhageners, it appears to be a manifestation of perturbations in his own theoretical perspective. The Copenhageners' projected on Einstein the classicism which was the source of instability and paradoxes in quantum mechanics, and Einstein projected on them his own demon—a concern about the status of the "real" which perturbed his own theoretical perspective.

Thus the source of one of the most important controversies in modern physics was the failure of both sides in the debate to build up on their innovations and create a consistent non-classical epistemology which would fully integrate the agency of the observer—arguably their biggest theoretical innovation. As a result, the symbolic systems that they constructed experienced internal perturbation. Rather than address the true source of these perturbations, both sides tried to alleviate the anxiety caused by the instabilities in their own theoretical perspective by projecting the source of the perturbation on the opponent. In their criticism, Bohr and the Copenhageners, who included a classical component in their own theory, tried to represent Einstein's position as "rigid adherence to classical theory," which completely disregarded very significant non-classical aspects of Einstein's theoretical heritage. Likewise, Einstein, whose own theoretical contributions brought forth the specter of subjectivity and self-referentiality, tended in his criticism to reduce the position of the Copenhageners to "anti-realism," which hardly reflected the complex character of quantum theory.

**Toward a Non-Classical Epistemology**

The debate between Einstein and the Copenhageners in many ways prefigured contemporary clashes between "realists" and "anti-realists" or "representationalists" and "anti-representationalists." The persistence of these and similar polarizations indicates that those involved in these debates are not critically addressing the real source of the controversy. As the above discussion suggests, there is only one way to resolve these disagreements and move forward: the construction of a genuine non-classical epistemology. But what features would characterize such epistemology? What will it look like?

The quantum debate was not an isolated phenomenon. What transpired in this debate was not related merely to physics or even science in general. The quantum controversy was an integral part of the overall evolution of Western thought and culture. The issues confronted by Einstein and the Copenhageners reflect the general trends of this evolution which supports the conclusions one can draw from this confrontation. In light of this evolution, the quantum debate, and the subsequent developments in physics, it is hard to imagine a new epistemology that would not recognize the agency of the knower. The recognition of agency inevitably involves the acceptance of the notion of the constructed nature of knowledge. Finally, as the analysis of the debate suggests, in order to resolve the paradoxes generated by the recognition of the agency of the observer, we must depart from classical dualism with its radical separation of the subject from the object. The only way out of the conundrum generated in the debate is to focus on the very process of construction which generates both the subject and the object.

Since the pioneering work of Jean Piaget on the construction of knowledge shifted the center of attention from the subject and the object to the process of construction,[54] a growing number of disciplinary and interdisciplinary approaches now focus on the process of construction. They include systems theory, communication theory, theory of emergence, constructionism, chaos theory, complexity theory, theory of autopoiesis, as well as more disciplinary fields such as evolutionary and developmental psychology, evolutionary and developmental biology, economics, management science, and others. Although there is still much to understand about the process of construction, several important features of this process can already be outlined.

As Piaget has shown, the process of construction involves operations which represent closed loops of action centered on their own functioning. Reflex is one example of such operation. In prehension, for instance, a neural signal and muscle contractions form such closed operational loop. The same loop is present in gustatory functions.[55] The very functioning of these operations conserves them. The more they function, the more they conserve themselves. Thus conservation is the most basic feature of the process of construction.

The functioning of operations requires stimulation. Something should activate the operation, trigger it into action. The triggering of the functional operation is obviously also an operation. Its function is to detect stimuli and send a signal that activates the main operation. It regulates the functional operation and helps to conserve it. Thus regulation is another important feature involved in the process of construction. The process of construction would be impossible without this vital combination of conservation and regulation.

Since the regulatory operation activates the main functional operation by reacting to external stimuli, it is capable of coupling the operation it regulates to other operations. The coupling brings individual operations into equilibrium with each other. The combination of operations is certainly more powerful than each individual operation in the coupling because it has a greater combinatorial power and can respond to a larger number of stimuli. For example, the combination of visual and auditory functions is activated by both visual and auditory stimuli. When infants combine these functions, they begin to see when they hear and hear when they see.[56] In this case, each function is stimulated not only by stimuli with which it is directly associated but also by those associated with the other function; each function is activated more often and therefore better conserved.

A coupling of operations has a much greater combinatorial power than each individual operation involved in it. A combination of operations is capable of functioning in an environment which is much richer than the one in which each of them functioned prior to their coupling.  Individual operations retain their identity while conserved in the coupling.  The interaction with a new and richer environment leads to a differentiation and the emergence of a variety of new derivatives of the coupling.  Their conservation requires equilibration and results in the creation of progressively more powerful structures which include many more operational possibilities.  A simple example illustrates this point.  Two separate operations A and B have only two possibilities each:  A and not-A, and B and not-B, for a total of four operations.  The combination of these operations offers sixteen possibilities.  With further equilibration of these derivatives, the number of possibilities increases exponentially to 256.

As one can see from the above description, there are two other features that define the process of construction.  Conservation of operations requires their equilibration.  This progressive equilibration creates more complex combinations and, as a result, a commensurate growth in disequilibrium.  Thus the progressive construction of more complex operations is characterized by a balance between equilibrium and disequilibrium.[57]

The process of construction is also characterized by emergence and downward causality.  In the current literature, emergence is defined as the appearance of new totalities which cannot be reduced to a mere sum of its components.  Indeed, the combination of operations offers much broader constructive possibilities.  It is more powerful than the mere agglomeration of the individual operations which compose it; it is a new and much richer totality.   It is worth noting that in this context the familiar dualism of continuity and discontinuity cease to be mutually exclusive.  Continuous process of equilibration results in what appears to be a discontinuity, or what is often also called "phase transition"--i.e., the creation of totalities qualitatively different from the operations which constitute them.

The term "downward causality" refers to the reversal of the causal sequence as it is recognized in the classical epistemology where the cause precedes the effect. The adaptation of the lower level operations to their more powerful combinations is a good example of such reversal. The construction of new combinations and, as a result, new environments have a profound effect on the original operations that constituted these combinations. They have to conform to the emerging totality which define them and limit their degrees of freedom. Thus the familiar relationship between cause and effect loses its classical linearity as the effect—the new totality—exerts powerful influence on its cause.

The above features certainly do not exhaust all there is to know about the process of construction. Much still has to be learned. As we continue to study the richness of structural possibilities at different levels of the organization of reality—from particles and atoms to cells and organisms, to symbolic operations and social evolution—we will certainly learn more details about the common features of this process which operate on all these levels.

The construction of knowledge is intimately related to the general process of construction that operates in the evolution of reality; it emerges from this general process. The operations that are performed in constructing knowledge are isomorphic to the processes that operate in the self-organization of reality. Therefore, in order to solve the problems involved in the creation of a truly non-classical epistemology, what we already know and will learn about the process of construction in general should be incorporated into how we know.

The problem of objectivity is one such problem. It was a major preoccupation at the time of the quantum debate and still continues to draw much attention from scientists and scholars.[58] The classical conception of objectivity revolves around traditional dualism with its unbridgeable gap between the subject and the object. In this conception, an objective description is one in which the subject has absolutely no impact on the object. Quantum mechanics challenged this conception by recognizing the need to include the observer in the act of observing. One of the innovations introduced by quantum theory was to acknowledge the essential connection between observers and their tools, on one hand, and objects, on the other. However, as a result of the vestigial survival of classical dualism in quantum epistemology, the new perspective has not resolved the problem of objectivity. This inconsistency, demonstrated in the above analysis of the quantum debate, has resulted in a paradoxical recognition of objects as ultimately inaccessible things-in-themselves.

There is a growing realization in the post-quantum world that objectivity requires an integration of the observer in the process of observing. In order to be objective, a description should involve observing the object and the observer (subject) at the same time. It is important to recognize that a physical system and an observer operate on very different levels of the self-organization of reality. They can be integrated in the same act of observing only on the basis of what they have in common. Otherwise the self-referential circularity of observing will only lead to what Nikolas Luhmann characterized as "infinite regress" since one must always ask "for the reasons behind the reasons."[59]

As has been stated earlier, the general process of self-organization underlies the evolution of reality, including our mental activity. The operations that we perform in constructing knowledge are isomorphic to the processes involved in self-organization. It is essential to keep in mind this fundamental connection between the evolution of reality and the evolution of our thought for resolving the problem of objectivity. The recognition of this connection leads to one important conclusion: knowledge production involves active construction, not passive observing.

Since conservation and regulation are integral to the process of the self-organization of reality, they are involved in every particular instance of this process. Therefore, in order to understand how a particular instance (object) of self-organization arises, it is important to identify the processes that are being conserved at this particular level of self-organization and the regulatory processes that are involved in their conservation. Also, since the process of self-organization of reality maintains a constant equilibrium between equilibrium and disequilibrium, any particular instance of this process should be viewed in the context of such equilibrium, that is, as a critical juncture between the processes that are being equilibrated and the kind of disequilibria that this equilibration generates. Observing these conditions is essential for establishing non-classical objectivity of a description of reality.

We are only beginning to understand how reality organizes itself and how knowledge, which is one of the products of this self-organization, is constructed. There is certainly much more to learn, and it is important to pursue this knowledge. Only through an understanding of self-organization, objective knowledge of reality can be achieved: that is, the kind of knowledge that fully integrates the knower into the process of construction of knowledge, in contrast to classical knowledge where the knower is located on the transcendent plane and the foundational propositions, or axioms, on which knowledge is constructed lie outside the field of observing.

Our civilization has already entered a stage in its evolution when machines, computers, and robots largely free us from repetitive physical or metal labor and will probably completely replace humans in this capacity.  Increasingly, our civilization puts premium on the production of knowledge and the creation of new ideas.  How much more productive can we be if we control our own creativity?  If we are capable of avoiding futile conflicts, similar to the one which divided the two of the greatest minds of the 20th century and polarized the scientific community?  What could have resulted from their cooperation?  How many new ideas would have been born?  We will never know.

An understanding of the process of construction will allow us to control better our own creativity.  By mastering this process we can learn to be creative when we want to, not only when we can.  We can make production of knowledge a much more efficient, more cooperative and more orderly process, rather than turn it into a largely futile and wasteful exercise of power.